\documentclass[aps,prb,reprint]{revtex4-2}
\usepackage{epsfig}
\usepackage{dcolumn}
\usepackage{physics}
\usepackage{graphicx,graphics}
\usepackage{amsmath}
\usepackage{amssymb}
\usepackage{bm}
\usepackage{color}
\usepackage[utf8]{inputenc}
\usepackage[colorlinks,linkcolor={blue},citecolor={blue},urlcolor={blue}]{hyperref}
\usepackage{mathtools}
\usepackage{booktabs}
\usepackage{float}

\newcommand{\bk}{{\bm k}}
\begin{document}
  \title{Impurity States and Indirect Exchange Interaction in Irradiated Graphene}

\author{Modi Ke}
\author{Wang-Kong Tse}
\affiliation{Department of Physics and Astronomy, The University of Alabama, Tuscaloosa, AL 35487, USA }

\begin{abstract}
We theoretically investigate the impurity levels and exchange interaction between magnetic impurities in graphene driven by an off-resonant circularly polarized light field.
Our analysis captures the non-perturbative effects resulting from scattering with magnetic impurities with a strong onsite potential. Under irradiation, a dynamical band gap opens up at the Dirac point, allowing impurity levels to exist inside the gap. These impurity levels are shown to give rise to a resonance feature in the exchange energy for impurities located either at the same or different sublattices. The exchange interaction also shows a wider spatial range of antiferromagnetic behavior due to irradiation. Our work demonstrates that the exchange energy of magnetic impurities in graphene is extensively tunable by light irradiation in the presence of strong potential scattering.
\end{abstract}

\maketitle
\section{Introduction}\label{Intro}

Magnetic two-dimensional (2D) materials have recently attracted considerable attention \cite{park_rev,burch_rev}. 
In particular, there has been much interest over the years in graphene with magnetic degrees of freedom that may host  interesting spintronic phenomena and applications \cite{han_rev,avsar_rev}. Although graphene is not intrinsically magnetic, magnetism in graphene can arise extrinsically from adatom deposition, vacancies and edge termination \cite{yazyev}. In the presence of
a dilute concentration of magnetic impurities, the dominant mechanism of exchange interaction between the impurity spins is the long-range indirect exchange interaction known as the Ruderman–Kittel–Kasuya–Yosida (RKKY) interaction, which is an effective spin-spin interaction mediated by the itinerant electrons in a non-magnetic host material. Thus, RKKY interaction  plays an important role in understanding the magnetic properties of graphene with a dilute concentration of magnetic impurities \cite{PowerCrystal,min2017}.
The RKKY interaction in intrinsic  graphene can be ferromagnetic or antiferromagnetic depending on the positions of the two magnetic impurities on the lattice, exhibiting a faster spatial decay $\sim 1/R^3$  than in regular 2D electron systems \cite{serami,blackshaffer,Satpathy1}. One important manifestation of graphene valley physics is the presence of short-range oscillations arising from intervalley scattering when the impurities are in the zigzag direction of the honeycomb lattice. These short-range oscillations are known to be present generically in any multi-valley electron system \cite{RKKYvalleys1} for which graphene serves as a specific example.
In extrinsic graphene where free carriers are present in the conduction or valence band, the RKKY interaction exhibits the typical Friedel oscillations with a half-Fermi wavelength period, in addition to possible valley-induced short-range oscillations occurring along the zigzag direction \cite{Satpathy2}.

The ability to control magnetic properties using an external field  offers a promising route to realize new material behaviors and functionalities.
In dilute magnetically doped systems, controlling the indirect exchange interaction could provide a path towards manipulation of the magnetic ordering of the impurity spins.
A number of theoretical works have addressed the exchange interaction in irradiated systems \cite{Lightexchange1,Lightexchange2,Lightexchange3,Lightexchange4,Lightexchange5} and demonstrated that the exchange coupling sensitively depend on the parameters of the driving light field.  In graphene in particular, 
off-resonant circularly polarized irradiation can strongly modify the sign and magnitude of the RKKY interaction between magnetic impurities due to the opening of light-induced dynamical gap at the Dirac point ~\cite{Modi1}. It is not clear, however, whether this remarkable effect remains robust for impurities that are characterized by strong potential scattering. Examples of such impurities are those with a large onsite potential including vacancies \cite{peres2006,nanda2012} (where theoretically the potential becomes infinite) or adatoms that couple resonantly with the host electrons (such as hydrogen adatoms in  graphene~\cite{Hydrogen1,Hydrogen2,Hydrogen3}). In these systems, repeated potential scattering processes between the host electrons and the impurities are important, which give rise to the formation of resonant impurity levels. 

Generally, the interaction between electrons and a magnetic impurity can be characterized by
both its spin-independent and spin-dependent contributions. 
The RKKY coupling is usually calculated theoretically considering  the second-order perturbation due to the exchange coupling between the host electrons and the impurity spins.
The role of the  spin-independent potential scattering contribution is neglected which is justified when the potential scattering strength is weak.
In contrast, magnetic impurities with a large onsite potential interaction causes strong electron-impurity scattering and the usual RKKY-type perturbation theory for the indirect exchange interaction  breaks down. 
To capture the effects of strong potential scattering in these systems, one would need to go beyond the standard RKKY treatment and include the effects of electron-impurity scattering to all orders of the scattering strength.  

In this work, we consider dilute magnetically doped graphene under electromagnetic irradiation 
and elucidate the consequences of strong potential scattering on the indirect exchange interaction between the impurity spins. The purpose of this work is to study how irradiation can modulate the impurity energy levels arising in the strong potential scattering regime, and investigate how  strong potential scattering
modifies the effects of irradiation on the indirect exchange interaction. To address these questions,  we formulate a non-perturbative theory that goes beyond the typical RKKY treatment to calculate the resonant impurity levels under irradiation and the  non-equilibrium indirect exchange interaction between the magnetic impurities.

The rest of the paper is organized as follows. In Sec.~\ref{Formulation} we present our theoretical model of graphene with two impurity spins under circularly polarized light. We focus on weak driving fields and present the corresponding Floquet Hamiltonian and Green's functions applicable in this regime. Using Floquet-Keldysh formalism, we then derive a general formula for the time-averaged interaction energy between two impurity spins in irradiated graphene. In Sec.~\ref{Potential Impurities}, we introduce the impurity model
and calculate the impurity levels in the presence of two impurities.  Numerical results for the time-averaged exchange energy is then presented in Sec.~\ref{numerics}. Sec.~\ref{concl} concludes the paper with a summary of our main findings.

\section{Formulation}\label{Formulation}

\subsection{Setup}
The low-energy electrons in single layer-graphene are described by the following continuum Dirac Hamiltonian near the $K$ and $K'$ points of the Brillouin zone,
\begin{subequations}\label{h0tau}
\begin{eqnarray}
{H}_0 &=& \int d\bm{r} \Psi^{\dagger}(\bm{r})\mathcal{H}_0\Psi(\bm{r}),\label{h0taua} \\
\mathcal{H}_0 &=&  -i \hbar v_F(\tau \sigma_x \partial_x+ \sigma_y \partial_y),\label{h0taub}
\end{eqnarray}
\end{subequations}
where $\Psi,\Psi^{\dagger}$ are the annihilation and creation field operators, $v_{F}=10^{6}$ ms$^{-1}$ is the band velocity and $\tau =1$ $(\tau=-1)$ corresponds to the $K$ $(K')$ Dirac points. $v_{F}$ is related to the graphene tight-binding model parameters by $\hbar v_{F}=3\Lambda a/2=6.6\;\mathrm{eV\AA}$ with $\Lambda =3$ eV being the nearest-neighbor hopping amplitude and $a=1.4$ \AA\; the carbon-carbon distance. We model the interaction of the graphene electrons with a magnetic impurity by a short-range interaction $V(\bm{r})$ between the electron and  impurity, and the total Hamiltonian is given by
\begin{eqnarray}\label{Hamilt1}
\mathcal{H}=\mathcal{H}_0 + V(\bm{r}). \label{H0V}
\end{eqnarray}

We consider a circularly polarized (CP) light field ${\bm E}=E_0[\cos{(\Omega t)}\hat x+\sin{(\Omega t)}\hat y]/\sqrt{2}$ normally incident on the graphene plane with a frequency $\Omega$ and a field amplitude  $E_{0}$. The field couples to the Hamiltonian Eq.~\eqref{h0tau} via the minimal coupling scheme and the resulting time-dependent Hamiltonian of the irradiated system $\mathcal{H} = \mathcal{H}_0(t)+V(\bm{r})$ becomes time-periodic with
\begin{eqnarray}\label{tdph}
  \mathcal{H}_0(t) &=& -i\hbar v_F (\tau \sigma_x \partial_x+\sigma_y \partial_y)\nonumber
\\
&&+A [-\tau \sigma_x \sin{(\Omega t)}+\sigma_y \cos{(\Omega t)}],
\end{eqnarray}
where we have defined $A= v_F e E_0/(\sqrt{2}\Omega)$ as the driving amplitude. In this work, we are interested in high frequency driving with $\hbar\Omega$  exceeding the electronic bandwidth $6\Lambda$, and low drive amplitudes $A$ such that the dimensionless driving strength  $\mathcal{A}=A/(\hbar \Omega)\ll 1$. In this weak drive regime, the irradiation is off-resonance. After the initial switching on of the laser and the initial transients have been washed out, the system settles into a  non-equilibrium steady state (NESS). The driven system is then periodic in time, and it is convenient to work with the Floquet picture. As detailed in Refs.~\cite{busl,Modi1},
a suitable unitary transformation $U$ exists that makes explicit the contributions from processes  due to different photon numbers in the transformed Floquet Hamiltonian  $\tilde{\mathcal{H}}_{F} = U^{\dagger}\mathcal{H}_{F}U$. This is called the photon number representation and allows for a systematic accounting of processes due to virtual photons, one-photon resonance, two-photon resonance  and so on. In our currently considered  high-frequency, off-resonance regime, only virtual photon processes are relevant and one can neglect contributions due to one-photon and higher-order resonances. This approximation is called F$_0$ approximation ~\cite{busl} and the resulting  Floquet Hamiltonian takes a  block-diagonal form $\tilde{\mathcal{H}}_{F} = \bigoplus_{m = -\infty}^{\infty}\tilde{h}_{m}$, where $\bigoplus$ stands for the matrix direct sum over the Floquet space, with the following $2\times 2$ block Hamiltonian $\tilde{h}_{m}$
\begin{eqnarray}\label{hf0}
\tilde{h}_{m}=F_0 \hbar v_F  \tilde{\bm{\sigma}}^*\cdot\bm{k}+\frac{\Delta}{2}\tilde{\sigma}_z+m \hbar  \Omega\mathbb{I}_{\tilde{\sigma}},
\end{eqnarray}
where $\Delta=\sqrt{ (2A)^2+(\hbar\Omega)^2}-\hbar \Omega$, $m\in \mathbb{Z}$ denotes the Floquet mode index, $\tilde{\bm{\sigma}}$ stands for the Pauli matrix vector in the photon number representation, and
\begin{eqnarray}\label{fN}
F_0 =\frac{1}{2} \left[1+\frac{\hbar \Omega}{\sqrt{(2A)^2+(\hbar \Omega)^2}}\right].
\end{eqnarray}
Eq.~\eqref{hf0} describes a quasienergy dispersion $\epsilon_{m,\bk}=\pm\sqrt{(F_{0}\hbar v_{F}k)^2+(\Delta/2)^2}+m\hbar\Omega$ of a ladder of gapped Dirac cones shifted by integer multiples of $\hbar  \Omega$, each with a renormalized band velocity $F_0 v_F$ and a photon-induced band gap $\Delta$ at the $K$ and $K'$ points. The retarded Green's function becomes
block diagonal
$\tilde{\mathcal{G}}^{{\rm R}} = \bigoplus_{m = -\infty}^{\infty}\mathfrak{g}^{{\rm R}}(\boldsymbol{k},\bar{\omega}-m\hbar\Omega)$ with $\mathfrak{g}^{{\rm R}}(\boldsymbol{k},\bar{\omega}-m\hbar\Omega)=[(\bar{\omega}+i \eta)\mathbb{I}_{\tilde\sigma}-\tilde{h}_{m}]^{-1}$
being the $2\times 2$ block Green's function corresponding to the $m^{\mathrm{th}}$ Floquet mode. The real-space representation of this Green's function at a single valley can be obtained from a Fourier transformation as  $\tilde{\mathcal{G}}^R(\bm{R},\bar{\omega})$
~\cite{Modi1}, where $\bm{R} = {\bm r}-\bm{r}'$.   The full real-space Green's function $\tilde{G}^R_0({\bm R},\bar{\omega})$ is then the sum of contributions from both valleys (detailed expressions of the real-space Green's functions are presented in  Appendix \ref{AppB}).

We now have the system's Floquet Hamiltonian that will allow us to use the Floquet Green's functions to derive other quantities of interest. In the following, we will derive an expression for the interaction energy between two impurities in irradiated graphene, which will allow us to calculate the exchange energy between two magnetic impurities in Sec.~\ref{numerics}.

\subsection{Interaction Energy}
To formulate a non-perturbative expression for the interaction  energy between the impurities, we start by calculating the  exact total energy $E$ of the electrons. Employing the familiar strategy~\cite{fetter2012quantum} of introducing a variable coupling constant $\gamma \in [0, 1]$, we consider an auxiliary problem with a Hamiltonian  having a scaled interaction $\gamma V$
\begin{eqnarray}\label{Hamilt2}
\mathcal{H}=\mathcal{H}_0 + \gamma V(\bm{r}), \label{H0gV}
\end{eqnarray}
with $\gamma=1$ corresponding to the Hamiltonian Eq.~\eqref{H0V} with impurities and $\gamma = 0$ to the case without impurities.
With the aid of this parameter $\gamma$ one obtains
\begin{eqnarray}\label{Eshift}
\frac{d}{d\gamma}E=\frac{1}{\gamma}\bra{\Psi_0(\gamma)}\gamma V\ket{\Psi_0(\gamma)}
\end{eqnarray}
where $\Psi_0(\gamma)$ is the exact eigenstate of the Hamiltonian Eq.~\eqref{H0gV}. The shift in total energy in the original problem due to interaction with the impurity potential $V$ can now be written as
\begin{eqnarray}\label{Eshift2}
\Delta E(t)= -i \int_{0}^{1}d\gamma {\rm Tr}\left\{\lim_{\substack{\boldsymbol{r}'\rightarrow\boldsymbol{r}\\ t'\rightarrow t}}V(\boldsymbol{r})G^<_{\gamma}(\boldsymbol{r},t,\boldsymbol{r}',t')\right\}, \label{delE}
\end{eqnarray}
where $G^<_{\gamma}$ is the lesser Green's function for the auxiliary problem with the scaled interaction $\gamma V$, and the trace is taken over the spin and sublattice indices as well as the position $\boldsymbol{r}$.

 In the following we will use the Floquet-Keldysh Green's function formalism and express all time-dependent quantities in the Floquet representation~\cite{Floquet}. Taking the time average of Eq.~\eqref{delE} we can  write the time-averaged shift in the total energy in the Floquet representation as
\begin{eqnarray}\label{Eshift3}
\overline{\Delta E}= -i \int_{0}^{1}d\gamma {\rm Tr}\left\{\int_{-\frac{\Omega}{2}}^{\frac{\Omega}{2}}\frac{d\bar{\omega}}{2\pi} V(\boldsymbol{r})\sum_{m} [G^<_{\gamma}(\boldsymbol{r},\boldsymbol{r},\bar{\omega})]_{m m}\right\},\nonumber\\
\end{eqnarray}
where $m \in \mathbb{Z}$ labels the Floquet  modes, and we use the symbol $\bar{\omega}$ to denote frequencies defined in the first reduced zone $(-\Omega/2,\Omega/2]$. We assume the irradiated graphene sheet is coupled to a  fermion bath that provides a thermalization mechanism through which energy relaxation takes place. The Floquet lesser Green's function is given by
\begin{eqnarray}\label{Glesser}
&&[G^<_{\gamma}(\boldsymbol{r},\boldsymbol{r},\bar{\omega})]_{m m}=
\\
&&\sum_{n n'}\int d\boldsymbol{r}'[G^R_{\gamma}(\boldsymbol{r},\boldsymbol{r}',\bar{\omega})]_{m n}[\Sigma^<(\bar{\omega})]_{n n'}[G^A_{\gamma}(\boldsymbol{r}',\boldsymbol{r},\bar{\omega})]_{n'm},\nonumber
\end{eqnarray}
where ${\Sigma}^{<} = 2i\eta \bigoplus_{n = -\infty}^{\infty} f(\bar{\omega}-n\hbar\Omega)\mathbb{I}_{\sigma}=2i\eta \mathcal{F}(\omega)$ is the lesser self energy due to bath coupling, with $\mathcal{F}(\omega)$ being the Fermi distribution function and $\eta$ is a phenomenological broadening parameter. In the above equation, the interacting retarded and advanced Green's functions $G^{\mathrm{R,A}}_{\gamma}$ can be expressed in terms of their non-interacting counterparts $G^{\mathrm{R,A}}_{0}$ using the Dyson's equation. Substituting Eq.~\eqref{Glesser} into Eq.~\eqref{Eshift3} and carrying out the integration over $\gamma$, after some algebra we find that Eq.~\eqref{Eshift3} under the F$_0$ approximation becomes
\begin{eqnarray} \label{DeltaEGV}
  \overline{\Delta E} &=& F_0\int_{-\frac{\Omega}{2}}^{\frac{\Omega}{2}}\frac{d\bar\omega}{\pi} \sum_{m} f(\bar\omega-m \hbar \Omega) \nonumber \\
                         &&\times{\rm Im}{\rm Tr}\{[\ln{(1-\tilde{G}_0^R F_0 V)}]_m\}.
\end{eqnarray}
We now focus on the scenario with two magnetic impurities labeled by L (left) and R (right), so that in the above equation $V$ will be the total potential energy $V = V_{\mathrm{L}}+V_{\mathrm{R}}$ of the two impurities.

In equilibrium, Eq.~\eqref{DeltaEGV} reduces to the so-called  Lloyd's formula~\cite{Lloyd}
\begin{eqnarray}\label{Lloydform}
\Delta E=\int_{-\infty}^{\infty}\frac{d\omega}{\pi} f(\omega){\rm Im}{\rm Tr}\{\ln{(1-{G}_0^R V)}\},
\end{eqnarray}
which has been used to compute the RKKY interaction between magnetic impurities~\cite{PowerCrystal}. We will not use Eq.~\eqref{DeltaEGV} since we are only interested in the interaction energy between the impurities and not the non-interacting part of the system's total energy.

Instead, we will rewrite Eq.~\eqref{DeltaEGV}  to single out specifically the  interaction energy, and for this purpose we will employ T-matrices. The electron Green's function for scattering with a single impurity potential $V$ can be written as $G^R = G_0^R+G_0^RT G_0^R$ where ${T} = V(1-{G}^R_{0}V)^{-1}$ is the T-matrix. The Green's function for scattering with two impurities is given by the Dyson's equation $G^R = G_0^R+G_0^R(V_L+V_R)G^R$, which, after some algebra, can be written in terms of the T-matrices ${T}_{L,R}$ for scattering with the single impurity potentials $V_{L,R}$
\begin{eqnarray} \label{2IGF}
{G}^R = {G}_0^R(1+ {T}_{R}{G}_0^R)(1-{T}_{L}{G}_0^R{T}_{R}{G}_0^R)^{-1}(1+{T}_{L}{G}_0^R).\nonumber\\
\end{eqnarray}
Eq.~\eqref{2IGF} nicely separates the scattering contributions with only one impurity $(1+ {T}_{L,R}{G}_0^R)$ and with both impurities $(1-{T}_{L}{G}_0^R{T}_{R}{G}_0^R)^{-1}$. This separation will be convenient  in the following when we extract the interaction energy shift between the two impurities.

Back to the expression of the exchange energy Eq.~\eqref{DeltaEGV}, we can use the Dyson's equation $1-{G}_0^R V ={G}_0^R ({G}^R)^{-1}$ expressed in the photon number basis to recast the argument inside the logarithm in terms of the full Green's function $(\tilde{G}^R)^{-1}$. Then, expressing Eq.~\eqref{2IGF} in the photon number basis to obtain $(\tilde{G}^R)^{-1}$ allows us to extract the interaction part of the energy shift as follows
\begin{eqnarray}\label{Exch}
\overline{\Delta E}= \frac{1}{\pi}\int_{-\frac{\Omega}{2}}^{\frac{\Omega}{2}}d\bar{\omega}\sum_{m} f(\bar\omega-m \hbar \Omega) F_0 \nonumber\\ \times{\rm Im} \Big\{{\rm Tr}[\ln{(1-\tilde{T}_L\tilde{G}_0^R\tilde{T}_R\tilde{G}_0^R)}]_m\Big\}.
\end{eqnarray}
Since $\tilde{G}_0^R$ is block diagonal, the  T-matrix can be resolved into $2\times 2$ blocks in the Floquet space:  $\tilde{T}_{L,R}=\bigoplus^{\infty}_{m=-\infty}\tilde{t}_{L,R}(\bar{\omega}-m\hbar\Omega)$, where  $\tilde{t}_{L,R} = F_0\tilde{V}_{L,R}(1-\tilde{g}^R F_0\tilde{V}_{L,R})^{-1}$
are the $2\times2$ T-matrices in the  photon number representation.
Converting the frequency variable from the reduced zone representation back to the extended zone with $\omega=\bar\omega-m \hbar \Omega$, Eq.~\eqref{Exch} then becomes
\begin{eqnarray}\label{eqint1}
&&\overline{\Delta E}= \frac{1}{\pi}\int_{-\infty}^{\infty}d\omega f({\omega}) F_0\\
&&\times{\rm Im} \Big\{{\rm Tr}\ln{[1-\tilde{t}_L(\omega)\tilde{g}^R(\boldsymbol{R},\omega)\tilde{t}_R(\omega)\tilde{g}^R(-\boldsymbol{R},\omega)]}\Big\},\nonumber
\end{eqnarray}
where in the extended zone, we have the T-matrix 
\begin{equation}
  \tilde{t}_{L,R}(\omega) = F_0\tilde{V}_{L,R}[1-\tilde{g}^R(0,\omega) F_0\tilde{V}_{L,R}]^{-1},
\end{equation}
and the Green's function 
%
\begin{eqnarray}
  \tilde{g}^R({\bm R},\omega)
  =e^{i {\bm K}\cdot {\bm R}}\sigma_{y}\mathfrak{g}^R({\bm R},\omega)\sigma_{y}+ e^{i {\bm K'}\cdot {\bm R}}\mathfrak{g}^R(\bm R,\omega). \nonumber \\ \label{tildeg} 
\end{eqnarray}
Eq.~\eqref{eqint1} is a central result of this paper, valid for arbitrary strengths of impurity scattering in irradiated graphene under a weak  high-frequency off-resonant Floquet drive. Appealingly, it has a similar form as the corresponding formula in equilibrium \cite{Abanin1,Mishchenko3,Mishchenko2}. 
In the following section, we specify the potential $V$ for the magnetic impurities in our calculations.

\section{Impurities with Magnetic and Potential Scattering}\label{Potential Impurities}

  \subsection{Impurity Model}
  The impurity potential $V$ generally consists of both spin-independent and spin-dependent scattering terms, which can be written as
\begin{eqnarray}\label{potential}
{V}_i\left(\boldsymbol{r}\right)=A_0\left(u
\mathbb{I}_{s}+\frac{\hbar\lambda}{2}\boldsymbol{s}\cdot\bm{S}_i\right)\delta\left(\boldsymbol{r}-\boldsymbol{R}_i\right)P_{\alpha},
\end{eqnarray}
where $A_0=3\sqrt{3}a^2/2$ is the area of unit cell, $i \in \left\{L,R\right\}$ labels the impurities at position  $\boldsymbol{R}_i$ (hereafter we set $\bm{R}_L$ as the origin and  $\bm{R}_R = \bm{R}$),  $u$ is the on-site potential energy, $\lambda$ is the exchange coupling strength, $(\mathbb{I}_{s},\boldsymbol{s})$ are the identity and Pauli matrices in the spin space, and $\bm{S}$ is the
angular momentum
of the localized impurity spin, which is taken to be 
along the out-of-plane direction $\bm{S} = S\hat{\bm{z}}$.
We mention in passing that such an impurity potential was also considered in Refs.~\cite{biswas2010,Mishchenko1,shiranzaei} where the effects of strong potential scattering in topological insulators and graphene were studied. We focus on the case where the impurities are either substitutional adatoms or vacancies so that they are located at either one of the two sublattices A or B.  Correspondingly $P_{\alpha}$ is the projection operator to sublattice $\alpha \in \left\{A,B\right\}$  with  $P_{A,B}=(\mathbb{I}_{{\sigma}}\pm \sigma_z)/2$, where $\mathbb{I}_{{\sigma}}$ is the identity matrix in the pseudospin space. The two impurities can be located at the same sublattice sites (AA) or different sublattice sites (AB), and separated along the zigzag or armchair directions.

For impurities with a  weak potential scattering  term $\vert u \vert \ll   \hbar\lambda\vert\mathbb{S}\vert/2$, $u$ can be  neglected and the resulting exchange interaction between two magnetic impurities reduces to the standard RKKY interaction that is perturbatively valid up to $\lambda^2$. In this work we are interested in the opposite limit in which the potential scattering is strong with $\vert u \vert \gg \hbar\lambda\vert\mathbb{S}\vert/2$ and cannot be ignored, necessitating the use of Eq.~\eqref{eqint1} for the calculation of the exchange energy. In the following we will discuss the formation of impurity levels and see that it plays an important role in the exchange energy in this regime.

\subsection{Impurity Levels}

The impurity energy levels are given by the real part of the poles of the full Green's function Eq.~\eqref{2IGF},
$\mathrm{det}(1-{T}_{L}{G}_0^R{T}_{R}{G}_0^R) = 0$.
In the photon number representation, since $\tilde{G}_0$ is block diagonal, the above  equation  simplifies to
$\mathrm{det}(1-\tilde{t}_{L}\tilde{g}^R\tilde{t}_{R}\tilde{g}^R) = 0$ for extended zone frequency $\omega$, giving
\begin{eqnarray}\label{EimpEq}
&&1-\mathcal{T}(\omega)\tilde{g}_{\alpha \beta}^R(\bm R,\omega)\mathcal{T}(\omega)\tilde{g}_{\beta \alpha}^R(-\bm R,\omega)=0,\nonumber\\
\end{eqnarray}
where
\begin{eqnarray} \label{Talpha}
\mathcal{T}(\omega)=\nu A_0  [1-\nu A_0\tilde{g}_{\alpha \alpha}^R(0,\omega)]^{-1},
\end{eqnarray}
is a T-matrix, and
\begin{equation} \label{effnu}
    \nu=F_0\left(u+\frac{\hbar \lambda}{2}s S\right),
\end{equation}
is an effective potential for up spins ($s = 1$) and down spins ($s = -1$). We note that $\mathcal{T}$ is independent of $\alpha$ following from the same property of $\tilde{g}_{\alpha \alpha}^R(0,\omega)$ (Eq.~\eqref{g0omega} in the Appendix~\ref{AppB}). Further insight can be gained by casting the  above condition into an  alternative form as $1-\nu A_0 G_{\rm eff,\pm}=0$, where $G_{\rm eff,\pm}$ is an effective Green's function
\begin{eqnarray}\label{geff}
G_{\rm eff,\pm} =\tilde{g}_{\alpha\alpha}^R(0,\omega)\pm \sqrt{\tilde{g}_{\alpha \beta}^R(\bm R,\omega)\tilde{g}_{\beta \alpha}^R(-\bm R,\omega)}.
\end{eqnarray}
The energies of the two impurities states are obtained from $A_0\mathrm{Re}G_{\rm eff,\pm} =  1/\nu$, with line broadenings given by $A_0\mathrm{Im}G_{\rm eff,\pm}$.
In equilibrium, there is no band gap and the impurity states have a finite broadening due to coupling to the continuum of graphene band states. A circularly polarized illumination opens up a dynamical gap at the Dirac point, and impurity levels falling inside  this gap are not broadened by coupling to band  states. 
\begin{figure}
 \begin{center}
            \includegraphics[width=1\columnwidth]{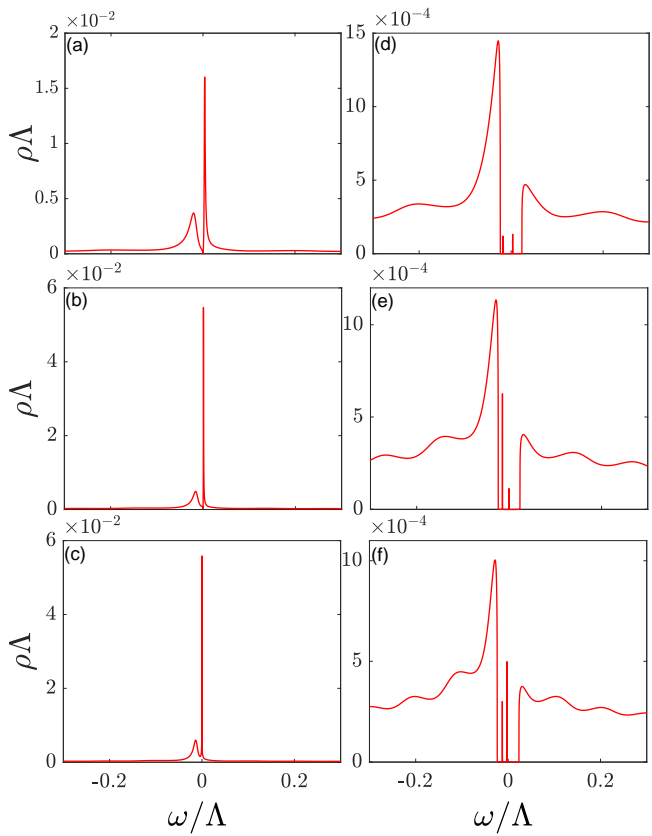}
            \end{center}
                \caption{Time-averaged local density of states of graphene evaluated at the impurity sublattice site A in equilibrium [panels (a)-(c)] and under irradiation with  $\mathcal{A}=0.06$ [panels (d)-(f)]. The  impurities are located at A and B separated along the armchair direction with  $u=100\Lambda$.
                  The upper two, middle two and lower two panels correspond to the values of impurity separation $R=31a,43a,55a$, respectively.
                }\label{LDOSTplot}
              \end{figure}
              
To examine the effects of irradiation, we first  calculate the time-averaged local density of states given by
\begin{eqnarray}\label{dos}
\rho(0,\omega)=-\frac{1}{\pi}\Im{{\rm Tr} [G^R(0,0,\bar\omega)]_{\alpha\alpha,m_{\omega}m_{\omega}}},
\end{eqnarray}
where $\alpha$ is the A/B sublattice of the impurity site at the origin and the trace is taken over the spin degrees of freedom. Given an extended-zone frequency value $\omega$ on the left-hand side of Eq.~\eqref{dos}, $\bar\omega$ and $m_{\omega}$ on the right-hand side can be found from $\omega=\bar\omega-m_{\omega}\hbar\Omega$ with  $m_{\omega}=-\rm{sgn}(\omega)\lceil{{|\omega|}/{\Omega}-{1}/{2}}\rceil$, where $\lceil\;\rceil$ is the ceiling function.
It should be noted that $G^R(0,0,\bar\omega)$ in Eq.~\eqref{dos} is the full interacting Green's function obtained from Eq.~\eqref{2IGF} in the presence of two impurities.  Throughout our numerical calculations in this section and in Sec.~\ref{numerics}, we  have chosen the following values:  frequency $\hbar \Omega=6.6\Lambda$, dimensionless driving strength $\mathcal{A} = A/(\hbar \Omega)=0.06$ and exchange coupling ${\hbar\lambda}|\bm{S}_i|/2 = 0.1\,\mathrm{eV}$. The frequency value is chosen to be larger than the electronic bandwidth $6\Lambda$ so that the driving field is off-resonant, and the driving strength is chosen so that it satisfies the weak drive condition  $\mathcal{A} \ll 1$. Fig.~\ref{LDOSTplot} shows, both in equilibrium and under irradiation, $\rho(0,\omega)$ for two impurities situated at A and B sublattice sites  and separated along the armchair direction. As expected, irradiation induces a gap in the local density of states and two sharp lines corresponding to the impurity levels emerge. 

\begin{figure}
 \begin{center}
            \includegraphics[width=1\columnwidth]{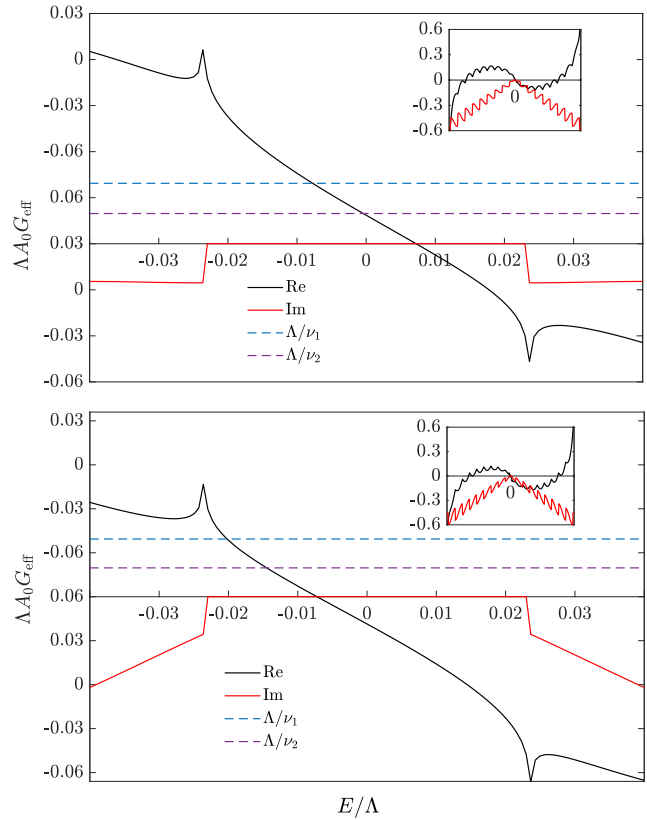}
            \end{center}
                \caption{Real and imaginary parts of the effective Green's function $G_{\rm eff,+}$ (upper  panel) and $G_{\rm eff,-}$ (lower  panel) for irradiated  graphene with $\mathcal{A}=0.06$ in the AB armchair case. The impurity  separation is $R=43a$ and effective  potentials are $\nu_1 = 50\Lambda$, $\nu_2 = 100\Lambda$.
                }\label{figc}
\end{figure}

We have examined the formation of the impurity levels for different impurity configurations (AA and AB, zigzag and armchair). For the AA case, both  impurity levels are found to be always below the Dirac point.  For the AB case however, the two impurity levels  remain below the Dirac point only for small $\nu$, and increasing $\nu$ eventually pushes the upper impurity level  above the Dirac point. These findings are  found to hold for both zigzag or armchair separations. Fig.~\ref{figc} illustrates the above for the AB armchair case through the graphical solution for $\omega_{{\rm imp}\pm}$. The upper panel demonstrates that how increasing $\nu$ can push the impurity level $\omega_{{\rm imp}+}$ pass zero to positive values, while the lower panel shows that $\omega_{{\rm imp}-}$ remains always negative. As expected, $\mathrm{Im}G_{\rm eff,\pm}$ is zero inside the band gap. We mention in passing that, in equilibrium, this ``Dirac point crossing'' behavior of $\omega_{{\rm imp}+}$ is associated with the transition from a repulsive to an attractive interaction between nonmagnetic impurities in graphene \cite{Mishchenko3}.
\begin{figure}
\begin{center}
            \includegraphics[width=1\columnwidth]{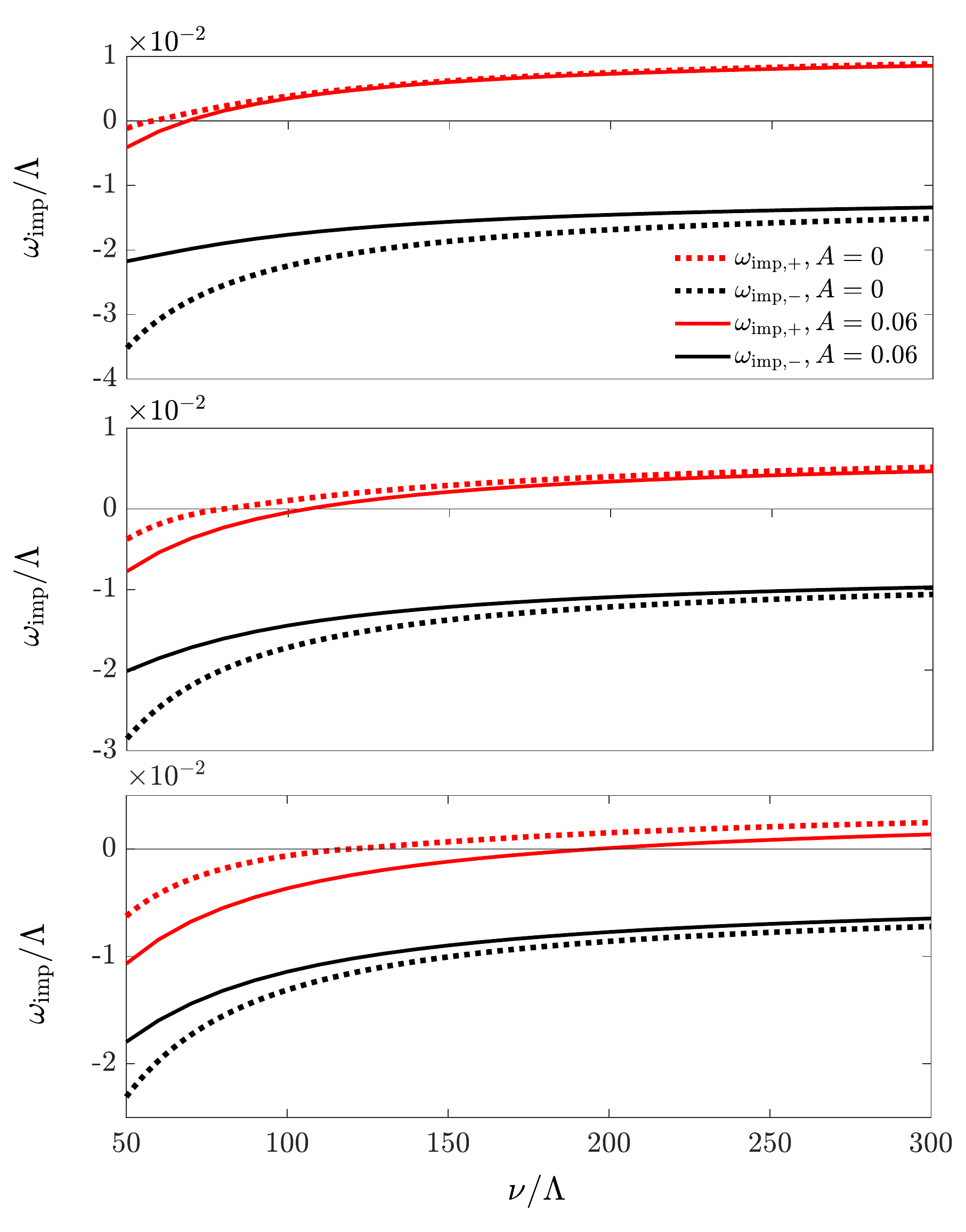}
            \end{center}
                \caption{Exact numerical result for  the impurity energy $E_{\rm imp}$ for graphene in equilibrium $\mathcal{A} = 0$ and irradiated with light $\mathcal{A}=0.06$ in the AB Armchair case as a function of the effective potential $\nu$. The upper, middle and lower panels correspond to impurity distances $R=31a,43a,64a$ respectively.
                }\label{figd}
\end{figure}

Fig.~\ref{figd} shows the numerically obtained impurity levels as a function of increasing potential $\nu$ under equilibrium and irradiated conditions. We see that the there is a considerable difference in the impurity energy levels between the two cases. In particular, turning on irradiation is seen to reduce the difference between the two impurity levels.
This difference is largest for small impurity strengths and gradually  decreases with $\nu$. Thus, for impurity strengths that are not too large, the impurity levels can be dynamically tuned by irradiation.
From Eq.~\eqref{EimpEq}, we have also derived approximate analytic results for the in-gap impurity levels
for large $\nu \gg \Lambda$ in both the AB and AA case.
We set the broadening parameter $\eta$ to zero in this calculation, so that  the impurity states inside the gap $\vert \omega_{\rm imp}\vert < \Delta/2$ are  undamped and have a zero linewidth. For the AB armchair case, we find
\begin{eqnarray}
  &&\omega_{{\rm imp},\pm}\approx\frac{\displaystyle \omega_0\ln{\left|\frac{2\Lambda}{\Delta}\right|}\pm|\sin{\theta_{AB}}|\xi K_1(\frac{\xi R}{F_0 \hbar v_F})}{\displaystyle\ln{\left|\frac{\Lambda}{\xi}\right|}},
\label{EimpEq4}
\end{eqnarray}
where $\theta_{AB}=\theta_{R}-\bm{K}\cdot \boldsymbol{R}$ with $\bm{K} = (2 \pi / (3\sqrt{3} a),0)$, {$\omega_0=-\pi \sqrt{3}F^2_0\Lambda^2/(2\nu \ln{(2\Lambda/\Delta)})$ and $\xi=\sqrt{(\Delta/2)^2-\omega_0^2}$}.

In the AA armchair case,
\begin{eqnarray}\label{EimpEq6}
&&\omega_{{\rm imp},\pm}\nonumber\\
&&\approx \frac{\displaystyle\omega_0\ln{\left|\frac{2\Lambda}{\Delta}\right|}\pm \sqrt{\omega_0^2\cos^2{\theta_{AA}}+(\frac{\Delta}{2})^2\sin^2{\theta_{AA}}} K_0(\frac{\xi R}{F_0\hbar v_F})}{\displaystyle\ln{\left|\frac{\Lambda}{\xi}\right|}}, \nonumber\\
\end{eqnarray}
{where $\theta_{AA}=\bm{K}\cdot \boldsymbol{R}$.}
\begin{figure}
 \begin{center}
            \includegraphics[width=1\columnwidth]{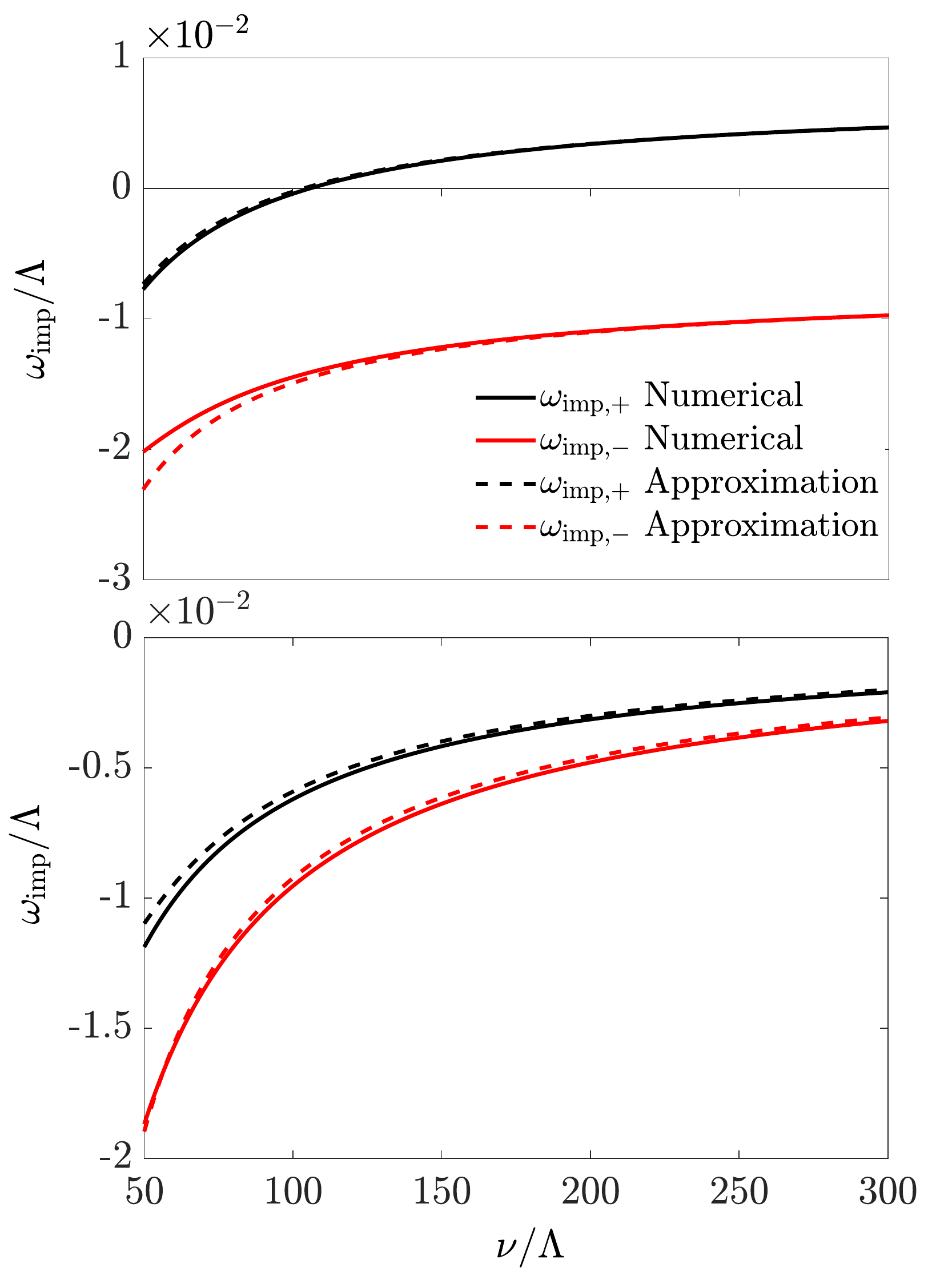}
            \end{center}
                \caption{Comparison of the exact numerical result and analytical approximation for the impurity energy $\omega_{\rm imp}$ for graphene irradiated with light $\mathcal{A}=0.06$ as a function of the effective potential $\nu$. The upper and lower panels correspond to the AB armchair case with  $R=43a$ and AA armchair case with $R=39a$.
                }\label{figEimpcomp}
\end{figure}
In Fig.~\ref{figEimpcomp} we plot the above approximate results Eq.~\eqref{EimpEq4} and Eq.~\eqref{EimpEq6} as well as the exact numerical results,
it is seen that they are in close agreement. Both solutions $\omega_{{\rm imp}\pm}$ are negative in the AA armchair case, whereas in the AB armchair case $\omega_{{\rm imp}+}$ turns positive for large enough $\nu$ (or small enough  $R$) with $\omega_{{\rm imp}-}$ remaining negative.

{Having investigated the impurity energy levels induced by the impurity potentials, we will present in the next section the results for the exchange energy. The physics of the impurity levels can be demonstrated by its effect on the exchange energy.}

\section{Exchange Energy}\label{numerics}

The important role of the impurity levels in the exchange energy can be recognized from the observation that the condition determining the impurity levels also appears directly in the integrand of Eq.~\eqref{eqint1}. This can be  appreciated more clearly if we take the exchange coupling $\lambda$ to be small and expand Eq.~\eqref{eqint1} up to leading order in $\lambda$. Then the time-averaged interaction energy takes the typical form $\overline{\Delta E} = J_{\alpha\beta}(\bm R)S_{L}S_{R}$ with an RKKY coupling strength  $J_{\alpha\beta}(\bm R)=({F_0^2\lambda^2\hbar^2}/{4})\chi_{\alpha\beta}(\bm R)$, where $\chi_{\alpha\beta}(\bm R)$ is the time-averaged spin susceptibility
\begin{eqnarray}\label{Susc}
&&\chi_{\alpha\beta}(\bm R)=-\frac{2F_{0}}{\pi} \int^{\infty}_{-\infty}d\omega f(\omega){\rm Im}\label{NLSS}
\\
&&\bigg\{\frac{A_0^2\tilde{g}_{\alpha \beta}^R(\bm R,\omega)\tilde{g}_{\beta \alpha}^R(-\bm R,\omega)}{\{[1-u_A\tilde{g}_{\alpha\alpha}^R(0,\omega)]^2-u_A^2\tilde{g}_{\alpha \beta}^R(\bm R,\omega)\tilde{g}_{\beta \alpha}^R(-\bm R,\omega)\}^2}\bigg\},\nonumber
\end{eqnarray} 
where $u_A=F_0 A_0 u$. In writing the above we have noticed that  $\tilde{g}_{\alpha\alpha}^R(0,\omega)$ is independent of the value of $\alpha$.
Eq.~\eqref{NLSS} generalizes the  so-called nonlinear spin susceptibility studied in  Ref.~\cite{NonlinearSus} to the non-equilibrium regime. The vanishing of the denominator of the integrand in Eq.~\eqref{Susc} gives the impurity levels in the absence of exchange coupling. Further, in the limit when the onsite potential $u = 0$, it can be shown that Eq.~\eqref{Susc} gives precisely the RKKY interaction energy previously obtained in the perturbative limit \cite{Modi1}.

We now perform exact non-perturbative calculations for the exchange energy using Eq.~\eqref{eqint1}.
The delta function potentials in Eq.~\eqref{potential} allow the spatial integrations in Eq.~\eqref{eqint1} to be straightforwardly carried out. Numerical convergence of the frequency integration in Eq.~\eqref{eqint1} can be facilitated by rotating it from the real axis to the imaginary axis. To compute the exchange energy, we evaluate the interaction energies for impurity spins aligned in parallel and antiparallel along the $z$-direction and calculate their difference.  Eq.~\eqref{eqint1} then becomes
\begin{eqnarray}\label{eqint0}
&&\overline {\Delta E}_{\alpha \beta}({\bm R})=-\frac{F_{0}}{\pi}\sum_{s = \pm 1} \int^{\infty}_{0}d\omega \nonumber\\
&&{\rm Re}\Big\{ {\rm Tr}\ln\left[1-\mathcal{T}^{s}_{\alpha}\tilde{g}_{\alpha \beta}^R(\bm R,i \omega)\mathcal{T}^{s}_{\beta}\tilde{g}_{\beta\alpha}^R(-\bm R,i \omega)\right]\Big\},\end{eqnarray}
where
\begin{eqnarray} \label{Talpha}
\mathcal{T}^{s}_{\alpha}=\nu A_0  [1-\nu A_0\tilde{g}_{\alpha \alpha}^R(0,i \omega)]^{-1}.
\end{eqnarray}
For parallel spins configuration, $\mathcal{T}^{s}_{\beta}$ is given by the same expression in Eq.~\eqref{Talpha}; for antiparallel spins,  $\mathcal{T}^{s}_{\beta}$ is given by Eq.~\eqref{Talpha} with $\lambda$ inside $\nu$ replaced by $-\lambda$. The exchange energy is then obtained  ~\cite{blackshaffer,PowerCrystal} as $\overline{E}_{\alpha \beta}^{\mathrm{ex}} = \overline {\Delta E}_{\alpha \beta}^{\uparrow\uparrow}-\overline {\Delta E}_{\alpha \beta}^{\uparrow\downarrow}$, where $\overline {\Delta E}_{\alpha \beta}^{\uparrow\uparrow}$ and $\overline {\Delta E}_{\alpha \beta}^{\uparrow\downarrow}$ are the interaction energies for impurity spins aligned in parallel and antiparallel, respectively. 
\begin{figure}
 \begin{center}
            \includegraphics[width=1\columnwidth]{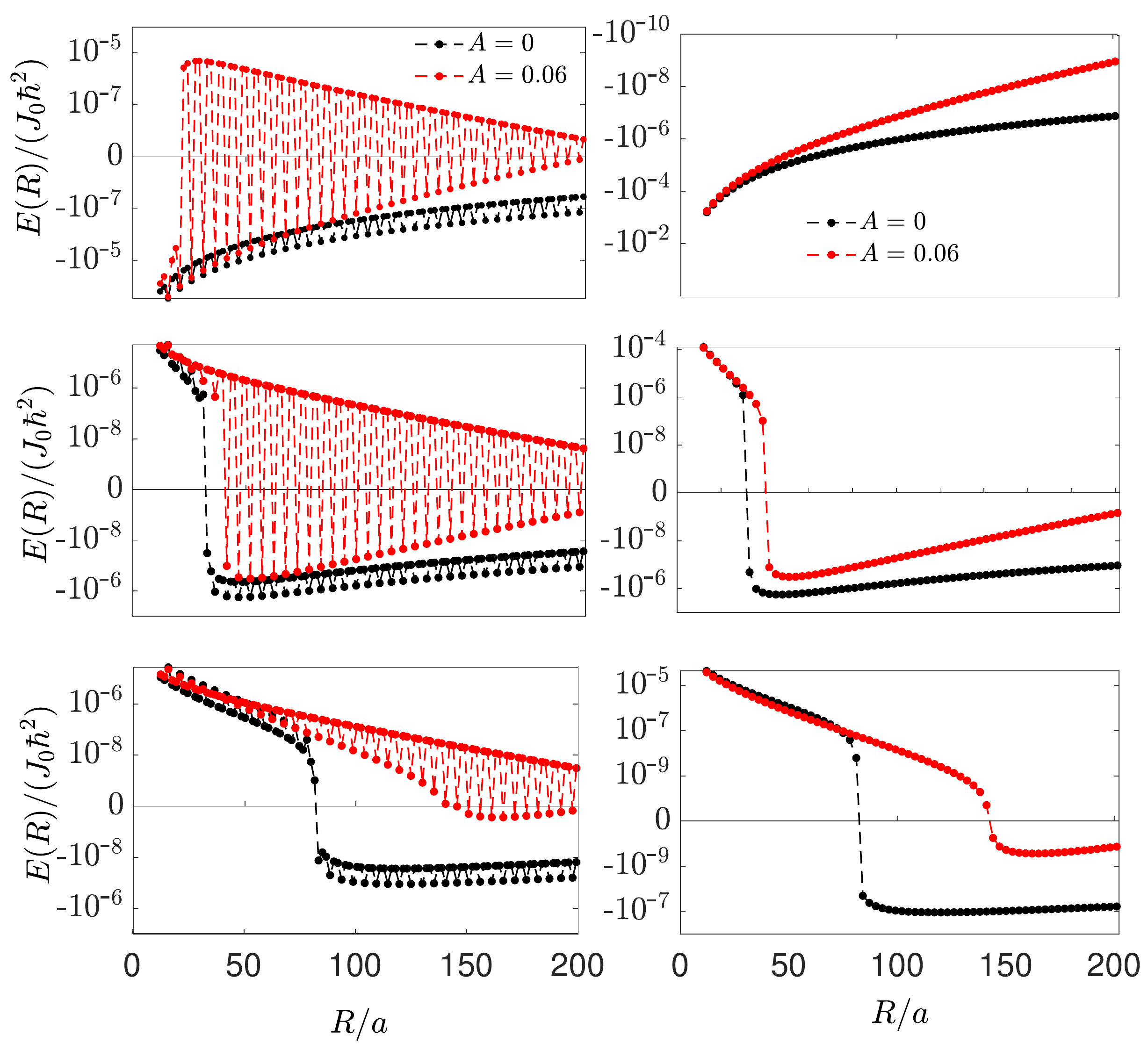}
            \end{center}
                \caption{Time-averaged exchange energy $E/(J_0 \hbar^2)$ versus impurity separation $R/a$ for impurities situated at sublattices AA and along the zigzag (left column) and armchair (right column) directions. Red (grey) line shows the equilibrium ($\mathcal{A} = 0$) case and black (dark) line shows the irradiated case ($\mathcal{A} = 0.06$). The upper, middle and lower panels correspond to onsite potentials $u/\Lambda=0,10,20$, respectively.}\label{fig1}
\end{figure}

\begin{figure}
 \begin{center}
            \includegraphics[width=1\columnwidth]{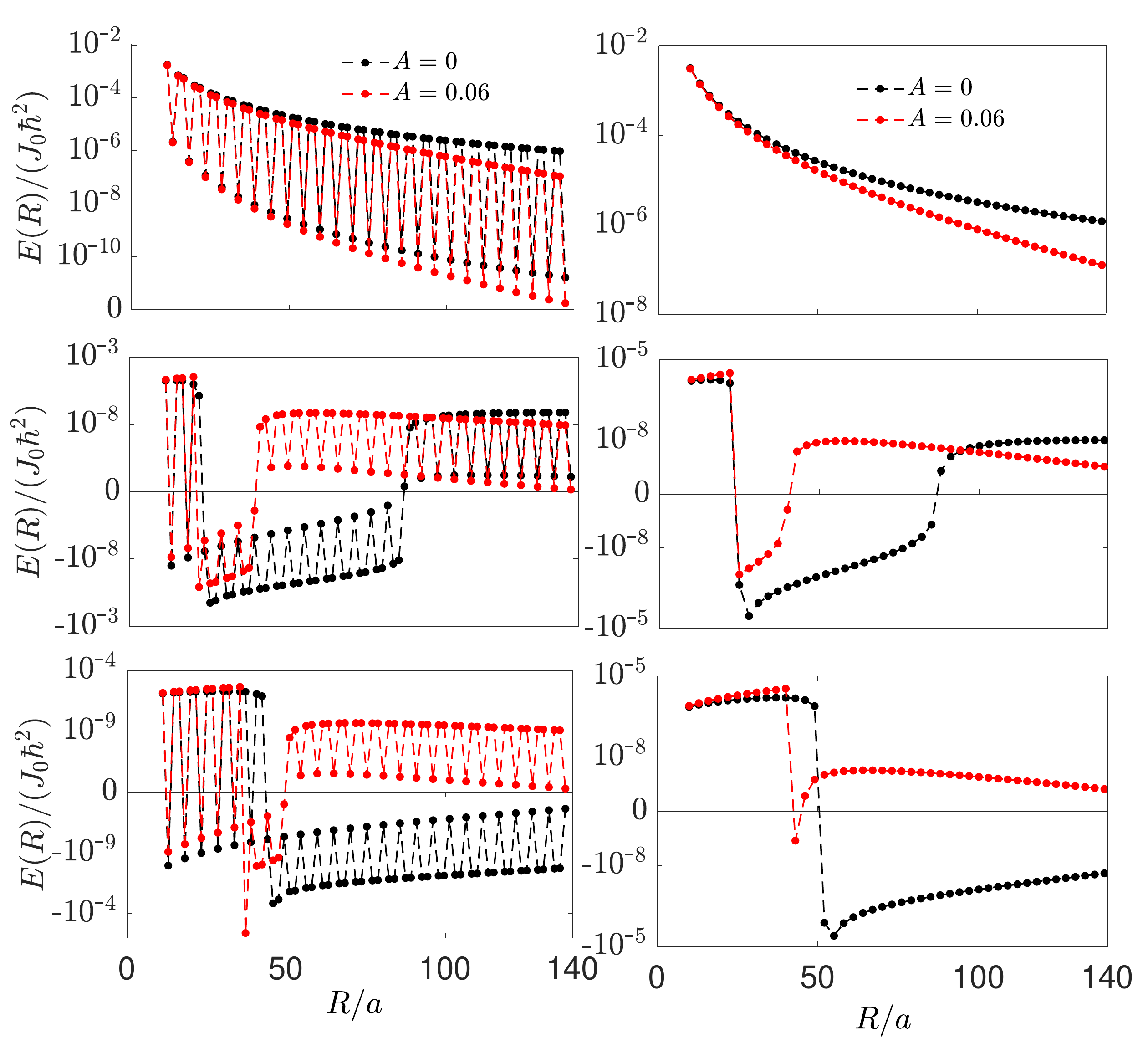}
            \end{center}
                \caption{Time-averaged exchange energy $E/(J_0 \hbar^2)$ versus impurity separation $R/a$ for impurities situated at sublattices AB and along the zigzag (left column) and armchair (right column) directions. Red (grey) line shows the equilibrium ($\mathcal{A} = 0$) case and black (dark) line shows the irradiated case ($\mathcal{A} = 0.06$). The upper, middle and lower panels correspond to onsite potentials $u/\Lambda=0,10,20$, respectively.}\label{fig2}
\end{figure}

Figs.~\ref{fig1}-\ref{fig2} show the evaluated  exchange energy in units of $J_0\hbar^2$ as a function of impurity separation for the AA and AB cases. $J_0$ is the characteristic scale for the exchange spin-spin coupling defined as
\begin{equation} \label{J0}
    J_0 = \frac{a}{2\pi \hbar v_F}\left(\frac{\lambda\hbar}{4a^2}\right)^2.
\end{equation}
We first focus on Fig.~\ref{fig1} showing the AA case for zigzag (left column) and armchair (right column) directions. The overall trend of the two sets of results are similar, and the main difference arises in the zigzag case where there are additional short-range oscillations due to intervalley scattering. For this reason we can focus ourselves first on the armchair case (right column). The upper panel  shows the case with $u = 0$ in the absence of an onsite potential, with the negative values of the exchange energy indicating a ferromagnetic coupling. At a finite $u$ (middle panel), the exchange energy stays negative for large $R$ but changes sign to positive for small enough $R$. Thus, at large $R$ the exchange energy for finite $u$ behaves qualitatively in the same way as that for $u = 0$.
This sign change is  present in both equilibrium and irradiated cases and occurs at similar values of $R$.
A larger difference between the equilibrium and irradiated cases is observed at a larger $u$ (bottom panel). Here, the sign change in the irradiated case occurs at an $R$-value almost doubling that in the equilibrium cases. Thus one can see that irradiation has the effect of extending the region of antiferromagnetic exchange behavior over a wider range of impurity separation. A more dramatic difference is seen in the zigzag case (left column), as there are now short-range oscillations on top of the aforementioned behavior in the armchair case. One can see that irradiation increases the amplitudes of the short-range oscillations so that the exchange energy now periodically changes sign with a period given by $3\sqrt{3}a/2$. In the large $R$ regime under irradiation, while the armchair case shows a ferromagnetic exchange coupling, the zigzag case shows periodic positive values caused by these light-enhanced short-range oscillations. When $u$ increases (middle to lower panel), these oscillations shift the exchange energy to overall more positive values.

For the AB case (Fig.~\ref{fig2}), again we first focus on the armchair configuration (right column) where the exchange energy does not contain  short-range oscillations. The exchange energy favors  antiferromagnetic alignment of impurity spins when $u = 0$ (top panel). In the presence of a finite $u$ (middle and bottom panels), the exchange energy exhibits a wide dip extending to negative values at equilibrium. Under irradiation,  the range of this dip is reduced considerably, so that the exchange energy recovers a positive value at a smaller $R$. As we discuss below, this dip is a resonance feature induced by the emptying of the occupation of the upper impurity level. The zigzag configuration (left column) behaves in a similar way except with the short-range  oscillations superposed onto the overall trend of the exchange energy displayed by the armchair case.

\begin{figure}
 \begin{center}
            \includegraphics[width=1\columnwidth]{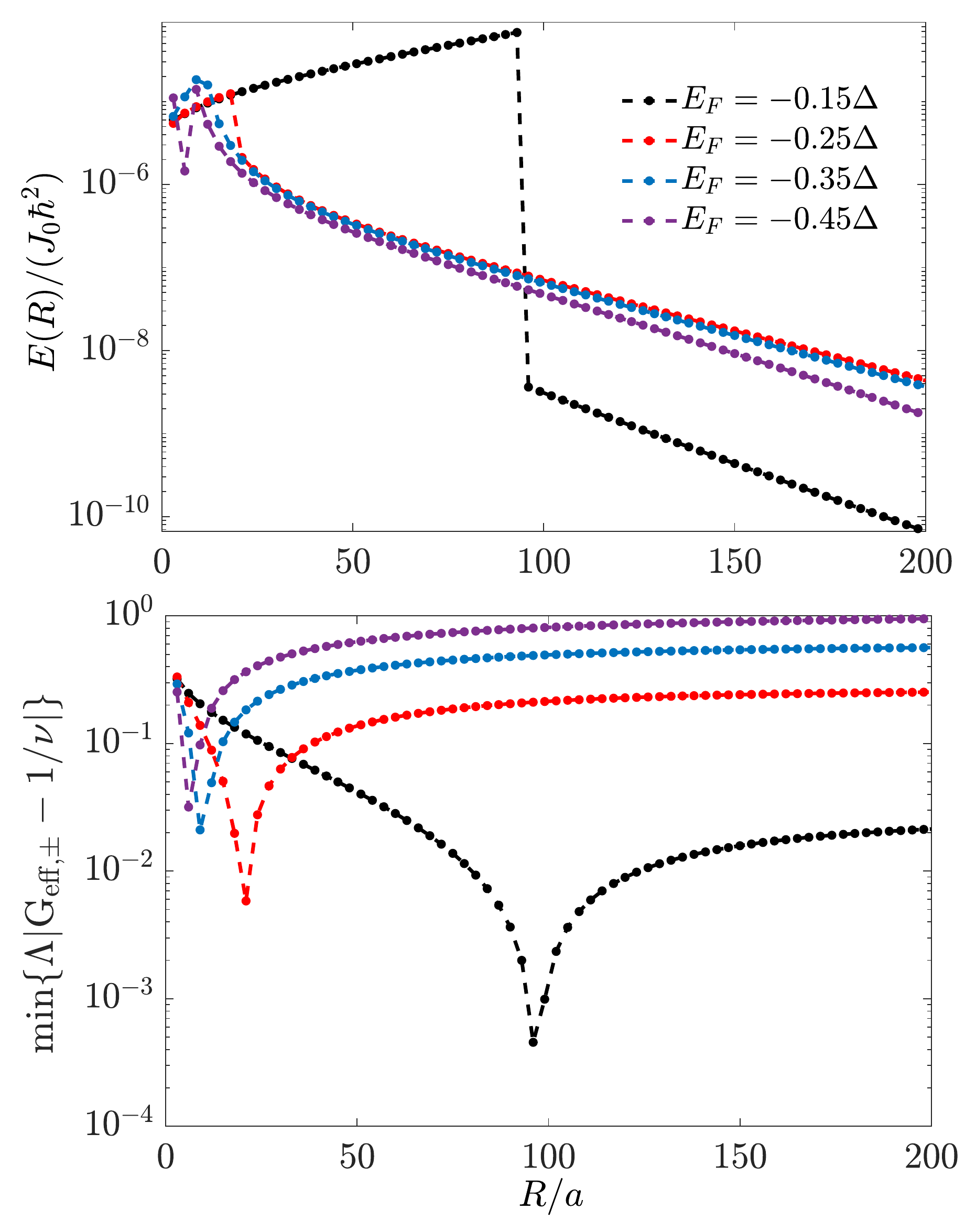}
            \end{center}
               \caption{The upper panel shows the time-averaged exchange energy $E/(J_0 \hbar^2)$ versus impurity separation $R/a$ for impurities situated at sublattices AA and along the  armchair direction with negative Fermi energy. The driving field strength is $\mathcal{A} = 0.06$. The lower panel plots the magnitude of the difference $G_{\rm eff,\pm}-1/\nu$, with its minimum giving the position of the resonance feature in the exchange energy.}\label{figAAEf}
             \end{figure}
             
In equilibrium, the appearance of a resonance feature in the AB case is associated with the upper impurity level crossing the Dirac point where the Fermi energy is located \cite{Mishchenko3}. Our  results in Fig.~\ref{fig2} shows that irradiation has the effect of narrowing the spatial range of this resonance feature. Furthermore, since irradiation induces a dynamical gap where undamped impurity levels can exist, this resonance feature is no longer restricted to occur only at $E_F = 0$ as in equilibrium, but can now occur for a range of Fermi energies within the gap. This can be demonstrated using the AA case as an example. In equilibrium, both impurity levels  always stay below the Dirac point and therefore there is no resonance feature due to crossing of the upper impurity level with the Fermi level located at the Dirac point. However, under a strong enough driving field both impurity levels can fall within the light-induced gap. Then, tuning the Fermi energy to a value in-between the two impurity levels will induce a resonance feature in the exchange energy. This is illustrated in  Fig.~\ref{figAAEf}, which shows the exchange energy for the AA armchair case at different nonzero Fermi energy values. The lower panel displays the locations of the resonant value of $R$ given by the minimum of each  curve. This value of $R$ is the closest physical distance on the graphene lattice that satisfies Eq.~\eqref{EimpEq}.
Comparing the bottom and the top panels shows that indeed the resonant $R$ gives rise to a corresponding resonant feature in the exchange energy displayed in the upper panel.
At $E_F = -0.15\Delta$, the exchange exhibits a large jump at $R = 96a$, similar to the resonance features in the AB armchair case shown in Fig.~\ref{fig2}. Hence, unlike in equilibrium where such a resonance feature only occurs in the AB case, irradiation can induce a resonance feature in both AA and AB cases.

A final remark on the tunability of the exchange energy by the driving field is in order. When the Fermi level is located at the middle of the light-induced gap, the case with no potential scattering ($u = 0$) offers a limited tunability on the exchange energy only through its magnitude. However, for impurities with strong potential scattering, the exchange energy becomes more extensively tunable
not only through its magnitude but also its sign. This can be seen, for example, in the AA armchair case in Fig.~\ref{fig1}. For $u = 0$, the exchange is ferromagnetic when the driving field is switched either off or on. In contrast, for $u = 20\Lambda$, there exists a range of $R \sim 80a\,-\,150a$ for which the sign of the exchange can be tuned from ferromagnetic to antiferromagnetic by switching on the driving field. Similarly, in the AB armchair case in Fig.~\ref{fig2}, the exchange remains antiferromagnetic for $u = 0$ in the presence or  absence of a driving field. In contrast, for $u = 20\Lambda$, there exists a range of $R \gtrsim 50a$ for which the sign of the exchange changes from ferromagnetic to antiferromagnetic under a driving field. Thus the presence of strong potential impurities enhances the tunability of the exchange interaction by the driving field and may present an interesting prospect for light-controlled switching of antiferromagnetic/ferromagnetic exchange coupling. 

\section{Conclusion} \label{concl}

We have presented a calculation of the impurity levels and exchange energy of magnetic impurities in a graphene sheet irradiated by a circularly polarized light in the weak drive regime. {By using the T-matrix and the Floquet Green's function formalisms, a theory of the exchange energy beyond the standard RKKY treatment is obtained, capturing the effects of a strong onsite potential and a weak driving light field. Due to opening of a light-induced dynamical gap at the Dirac point, impurity levels can exist inside the band gap.  We have obtained numerical and analytical results for the impurity energy levels inside the gap due to a pair of magnetic impurities. 
For impurities located at either the same (AA) or different (AB) sublattice sites,  irradiation has the effect of extending the spatial range of antiferromagnetic behavior in the presence of strong onsite potential, and the exchange energy exhibits a resonance feature that originates  from the crossing of the Fermi energy with one of the impurity levels. The presence of this feature in the same-sublattice case is unique under irradiation and is due to the induced band gap under circularly polarized radiation.}

\acknowledgments {This work was supported by the U.S. Department of Energy, Office of Science, Basic Energy Sciences under Early Career Award No. DE-SC0019326.}

\appendix{}

\section{Floquet Green's functions within the F$_0$ approximation} \label{AppB}

For completeness we include the expressions of the Floquet Green's functions in the F$_0$ approximation that were originally derived in Ref.~\cite{Modi1}.

The $2\times 2$ Hamiltonian $\tilde{h}_{n}$  [Eq.~\eqref{hf0}] in the photon number representation  gives the following real space Green's function after a Fourier transformation,
\begin{eqnarray}\label{pm1}
\mathfrak{g}^{\rm R}({\bm R},\omega) &=& \zeta \left\{\left[-(\omega+i\eta)\mathbb{I}_{\tilde\sigma}+\frac{\Delta}{2}\tilde{\sigma}_z\right]\chi_0(R,\omega) \right.\nonumber \\
&& \left.-\tilde{\bm{\sigma}}^*\cdot \hat{\bm{R}} \chi_1(R,\omega)\right\},
\end{eqnarray}
where $\zeta=[2\pi(\hbar v_F F_0)^2]^{-1}$,   $\hat{\bm{R}}$ is the unit vector along   ${\bm{R}}$. $\chi_0(R,\omega)$ and $\chi_1(R,\omega)$ are defined as
 \begin{eqnarray}
 \chi_0(R,\omega)=\frac{\pi i}{2}{H}^{(1)}_0\left[\kappa(\omega)R\right],
  \end{eqnarray}
  \begin{eqnarray}
 \chi_1(R,\omega)=-\frac{\pi}{2}  \hbar v_F F_0\kappa(\omega){H}_1^{(1)}\left[\kappa(\omega) R\right],
  \end{eqnarray}
 with  ${H}^{(1)}_0$ and ${H}^{(1)}_1$ being the zeroth-order and first-order Hankel functions of the first kind, respectively, and
   \begin{eqnarray}
 \kappa(\omega)=\frac{{\rm sgn}(\omega)}{\hbar v_F F_0}\sqrt{(\omega+ i\eta)^2-\frac{\Delta^2}{4}}.\label{pm3}
 \end{eqnarray}

The $2\times 2$ Green's functions at $\bm{R} = 0$ can be obtained  from a Fourier transformation of the momentum-space Green's  function with a momentum cutoff $\Lambda/\hbar v_F$,
\begin{eqnarray}
 &&\mathfrak{g}^R(0,\omega) =
 -\frac{\zeta }{2}\bigg(\omega \pm \frac{\Delta}{2}\bigg)\mathbb{I}_{\tilde\sigma}\left\{\ln{\bigg |\frac{\Lambda ^2}{\omega^2-({\Delta}/{2})^2}\bigg |}\right. \label{g0omega} \\
 &&\left.-i\pi F_0\frac{\Lambda }{\sqrt{\Lambda ^2+(\Delta/2)^2}}\bigg[\theta\bigg(\omega-\frac{\Delta}{2}\bigg)-\theta\bigg(-\omega-\frac{\Delta}{2}\bigg)\bigg]\right\}.\nonumber
 \end{eqnarray}
The full real-space Green's function consists of contributions from both valleys:
\begin{eqnarray}\label{FourG2}
 &&\tilde{G}^R_0({\bm R},\bar{\omega}) = e^{i {\bm K}\cdot {\bm R}} \Sigma_{y} \tilde{\mathcal{G}}^R({\bm R},\bar{\omega})\Sigma^{\dag}_{y} + e^{i {\bm K'}\cdot {\bm R}}\tilde{\mathcal{G}}^R({\bm R},\bar{\omega}),\nonumber\\
\end{eqnarray}
where $\Sigma_{y}=i\sigma_{y}\otimes\mathbb{I}_{\infty}$ is a unitary transformation between the bases for the transformed Floquet Hamiltonian at the $K$ and $K'$ valleys, and $\mathbb{I}_{\infty}$ is the identity matrix in the Floquet space. We also have $\tilde{\mathcal{G}}^{{\rm R}} = \bigoplus_{m = -\infty}^{\infty}\mathfrak{g}^{{\rm R}}(\boldsymbol{k},\bar{\omega}-m\hbar\Omega)$ with $\mathfrak{g}^{{\rm R}}(\boldsymbol{k},\bar{\omega}-m\hbar\Omega)=[(\bar{\omega}+i \eta)\mathbb{I}_{\sigma}-\tilde{h}_{m}]^{-1}$. The diagonal blocks of Eq.~\eqref{FourG2} takes the following form:
\begin{eqnarray}\label{GreenApp2}
&&[\tilde{G}_0^R({\bm R},\bar{\omega})]_m=\tilde{g}^R({\bm R},\bar{\omega}-m\hbar\Omega)=
\\
&&e^{i {\bm K}\cdot {\bm R}}\sigma_{y}\mathfrak{g}^R({\bm R},\bar{\omega}-m\hbar\Omega)\sigma_{y}+ e^{i {\bm K'}\cdot {\bm R}}\mathfrak{g}^R(\bm R,\bar{\omega}-m\hbar\Omega).\nonumber
\end{eqnarray}
\vfil\null

\bibliography{refs_str_1}
\end{document}